\documentclass[12pt,a4paper,final]{iopart}
\usepackage[breaklinks=true,colorlinks=true,linkcolor=blue,urlcolor=blue,citecolor=blue]{hyperref}
\usepackage{graphicx}
\usepackage{multirow}
\expandafter\let\csname equation*\endcsname\relax 
\expandafter\let\csname endequation*\endcsname\relax 
\usepackage{amsmath,amssymb,latexsym,MnSymbol}

\begin{document}

\title[Anomalous diffusion via fractional Brownian walks on a comb-like structure]{Investigating the interplay between mechanisms of anomalous diffusion via fractional Brownian walks on a comb-like structure}

\author{Haroldo V. Ribeiro$^{1,2,3}$, Angel A. Tateishi$^{4}$, Luiz G. A. Alves$^{1,3}$, Rafael S. Zola$^{1,2}$, Ervin K. Lenzi$^{1,3}$}
\address{$^1$Departamento de F\'isica, Universidade Estadual de Maring\'a, Maring\'a, PR 87020-900, Brazil}
\address{$^2$Departamento de F\'isica, Universidade Tecnol\'ogica Federal do Paran\'a - Apucarana, PR 86812-460, Brazil}
\address{$^3$National Institute of Science and Technology for Complex Systems, CNPq - Rio de Janeiro, RJ 22290-180, Brazil}
\address{$^4$Departamento de F\'isica, Universidade Tecnol\'ogica Federal do Paran\'a - Pato Branco, PR 85503-390, Brazil}
\eads{\mailto{hvr@dfi.uem.br}}

\begin{abstract}
The comb model is a simplified description for anomalous diffusion under geometric constraints. It represents particles spreading out in a two-dimensional space where the motions in the $x$-direction are allowed only when the $y$ coordinate of the particle is zero. Here, we propose an extension for the comb model via Langevin-like equations driven by fractional Gaussian noises (long-range correlated). By carrying out computer simulations, we show that the correlations in the $y-$direction affect the diffusive behavior in the $x-$direction in a non-trivial fashion, resulting in a quite rich diffusive scenario characterized by usual, superdiffusive or subdiffusive scaling of second moment in the $x-$direction. We further show that the long-range correlations affect the probability distribution of the particle positions in the $x-$direction, making their tails longer when noise in the $y-$direction is persistent and shorter for anti-persistent noise. Our model thus combines and allows the study/analysis of the interplay between different mechanisms of anomalous diffusion (geometric constraints and long-range correlations) and may find direct applications for describing diffusion in complex systems such as living cells.
\end{abstract}

\vspace{2pc}
\submitto{\NJP}

\section{Introduction}

Understanding diffusive motions is a long-standing problem in the physicists' agenda. Since the works of Einstein, Smoluchowski and Langevin, we know that one of the most striking patterns of classical (usual) free diffusion is the linear growth with time of the second moment of the particle positions. While this feature is ubiquitous in nature, many other patterns have been also observed for the second moment, both from analytical calculations and experimental data~\cite{Klafter}. These deviants behaviors are often called anomalous diffusion. 

Nowadays, efforts have been mainly focused on understanding the underlying physical mechanisms that lead to such deviations, since only the classification of the diffusive behaviors of a given system is not enough to fully characterize it. In this context, two common pathways for anomalous diffusion are long-range correlations in the particle positions and geometric constraints related with the structural complexity of the environment where the particles are moving. The former may be related to memory affects and an example is the fractional Brownian motion~\cite{Mandelbrot,Gripenberg}, where the second moment has a power-law dependence with time, that is, $\langle x^2(t)\rangle \propto t^{2h}$. For this case, the Hurst exponent $h$ classifies the anomalous diffusion in subdiffusion ($0<h<1/2$) or superdiffusion ($h>1/2$)~\cite{Bouchad,Metzler} and also recovers usual ($h=1/2$) and ballistic ($h=1$) diffusions as limiting cases. Anomalous diffusion due to geometric constraints is well exemplified by the comb model. In this model, diffusive particles walk on a two-dimensional space; however, motions in the $x$-direction are only allowed when the $y$ coordinate of the particle positions is zero. The result of this simple constraint is a subdiffusive motion (precisely, $\langle x^2(t)\rangle\sim t^{\alpha}$ with $\alpha=1/2$) and the appearance of a backbone (at $y=0$) and teeth (along the $y-$axis direction), which were originally proposed to mimic the quasilinear structure and dangling ends of percolation clusters~\cite{White,Ben-Avraham,Weiss}. 

Mainly due to its simplicity and ability of mimicking diffusive aspects of highly disordered systems, the comb model has been extensively studied and extended by means of analytical calculations~\cite{Baskin,Arkhincheev,Iomin0,Tateishi,Villamaina,Tateishi2,Lenzi} and also employed as simplified description of natural phenomena such as cancer proliferation~\cite{Iomin1,Iomin2}, transport of spiny dendrites~\cite{Mendez} and diffusion of ultra cold atoms~\cite{Iomin3,Sagi}. Moreover, as crowded environments tend to slow down the diffusive motion of particles by obstruction and/or trapping~\cite{Franosch}, the comb model could also be used as a toy model for describing diffusion in intracellular processes, where several researchers have reported the existence of subdiffusion and ergodicity breaking in the random walk of different biochemical compounds~\cite{Condamin,Unkel,Jeon,Burov,Leijnse,Sokolov,Cherstvy,Franosch,Brackley,Girst,Tabei,Cherstvy2} by using single particle tracking~\cite{Michaelis}. In the case of intracellular processes, the power-law exponents $\alpha$ describing the second moment ranges from $\alpha\approx0.1$ to $\alpha\approx0.9$ (evaluated both from ensemble and time average) and, in the case of ultra cold atoms, $\alpha$ appears as a function of lattice depth and assumes values larger than one ($\alpha>1$, superdiffusion). 

In addition to the anomalous scaling of the second moment, the work of Weiss~\cite{WeissM} has also shown that the diffusive processes of intracellular fluids may present long-range corrections as in the fractional Brownian motion, an empirical finding that have not been explored within the comb model framework. Also, recently, Yamamoto \textit{et al.}~\cite{Yamamoto} have provided evidence (from molecular dynamics simulation) that the anomalous diffusion of water molecules on cell membrane surface arises from both divergent mean trapping time and long-range correlated noises. Here, we extend the comb model by considering a Langevin-like equation where the noises are long-range correlated. Our model thus combines two of the main mechanisms of anomalous diffusion: geometric constraints and long-range correlations, allowing a more complete description of complex diffusive motions and also the study of the interplay between these two mechanisms. By carrying out computer simulations, we show that the correlations in the $y-$direction affect the diffusive behavior in the $x-$direction in a non-trivial fashion and that a quite rich diffusive scenario emerges from these long-range correlations. We report that diffusive process in the $x-$direction of this generalized version of the comb model can be usual, superdiffusive or subdiffusive, depending on both corrections in the $x$ and $y$ directions. We further show that the long-range correlations affect the probability distribution of the particle positions in the $x-$direction: the distribution becomes more leptokurtic when the noise in the $y-$direction is persistent whereas, for anti-persistent noise, it becomes less leptokurtic and approaches the Gaussian distribution as a limiting case.

\section{Generalized comb and long-range correlations}

We start by writing the generalized diffusion equation describing the motion under the comb structure~\cite{Arkhincheev2}:
\begin{equation}\label{eqcombusual}
\frac{\partial}{\partial t} \rho(x,y;t) = \delta(y) D_x\frac{\partial^2}{\partial x^2} \rho(x,y;t) + D_y\frac{\partial^2}{\partial y^2} \rho(x,y;t)\,,
\end{equation}
here $\rho(x,y;t)$ is the joint probability of the particle positions ($x$ and $y$ coordinates) as a function of time $t$, $D_x$ and $D_y$ are the diffusive coefficients in the $x$ and $y$ directions. Note the presence of the Dirac delta $\delta(y)$ multiplying the spatial derivative with respect to $x$ and, consequently, limiting the diffusion in the $x-$direction only for $y=0$. In order to investigate the role of long-range correlations in this diffusive process, we propose to numerically solve the following coupled Langevin-like equations
\begin{eqnarray}\label{leqcomb}
x(t+dt) &=& x(t) + \beta_x \, \delta(y) \eta_x(t) \nonumber\\
y(t+dt) &=& y(t) + \beta_y \, \eta_y(t)\,,
\end{eqnarray}
where $\beta_x$ and $\beta_y$ are constants related to $D_x$ and $D_y$, $\eta_x(t)$ and $\eta_y(t)$ are fractional Gaussian noises~\cite{Mandelbrot,Gripenberg}, both with zero mean ($\langle\eta_x(t)\rangle = 0$ and $\langle\eta_y(t)\rangle = 0$), unitary variances ($\langle\eta_x^2(t)\rangle = 1$ and $\langle\eta_y^2(t)\rangle = 1$) and correlations functions (for $t_1\neq t_2$) given by 
\begin{eqnarray}\label{eqcor}
\langle \eta_x(t_1)\, \eta_x(t_2) \rangle &\sim& h_x (2h_x-1)|t_1-t_2|^{2h_x-1}~~\text{and}\nonumber \\
\langle \eta_y(t_1)\, \eta_y(t_2) \rangle &\sim& h_y (2h_y-1)|t_1-t_2|^{2h_y-1}\,.
\end{eqnarray}
Notice that the noises $\eta_x(t)$ and $\eta_y(t)$ are power-law correlated with scaling exponents $(2h_x-1)$ and $(2h_y-1)$, where $h_x$ and $h_y$ are the so-called Hurst exponents. For $h_{x,y}<1/2$, the correlation functions have negative signs and the noises are anti-persistent, meaning that positive values are followed by negative values (or vice-versa) more frequently than by chance; while for $h_{x,y}>1/2$ the correlations are positive, meaning that positive values are followed by positive values and negative values are followed by negative values more frequently than by chance. Fully persistent noises occur as a limiting case when $h_{x,y}=1$, whereas normal Brownian (white) noises correspond to the limit of $h_{x,y}=1/2$.

\begin{figure}[!ht]
\centering
\includegraphics[scale=0.5]{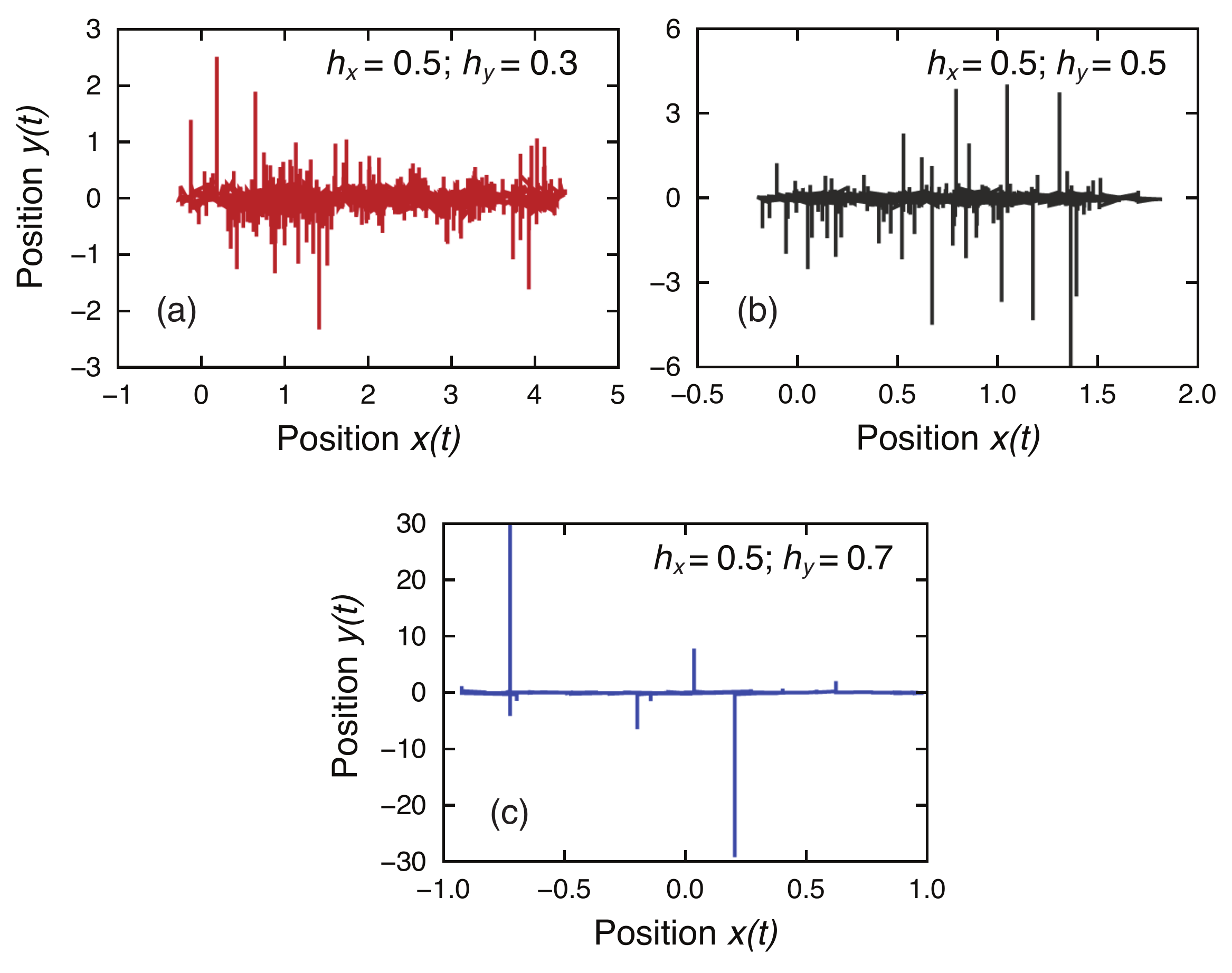}
\caption{(color online) {Simulation of the comb model via a Langevin equation.} Examples of simulated trajectories from the generalized comb model described by Eq.~\ref{leqcomb} with $\varepsilon=\beta_x=\beta_y=0.1$. We have kept $h_x=0.5$ and considered three values of the Hurst exponent $h_y$: (a) $h_y=0.3$, (b) $h_y=0.5$ and (c) $h_y=0.7$. Note that the walker returns less often to $y\approx0$ when the noise is persistent in the $y$-direction ($h_y>0.5$); whereas, for an anti-persistent noise ($h_y<0.5$), the walker returns more often to $y \approx 0$ when compared with the usual comb ($h_x=h_y=0.5$).}\label{fig1}
\end{figure}

When running out the simulations of the Langevin equations, we have considered a narrow band of thickness $\varepsilon$ along the $x-$axis, inside which the diffusion in the $x-$direction occurs. This strip mimics the effect of the Dirac delta $\delta(y)$ (that appears multiplying the noise $\eta_x(t)$ in Eq.~\ref{leqcomb}) and we have verified that the value $\varepsilon$ has no influence in the diffusive process, as long as $\varepsilon$ and noises amplitudes $\beta_x$ and $\beta_y$ are of the same order of magnitude. This restriction ensures that the variable $y$ presents no dynamics inside the strip. For our proposes, we have fixed  $\varepsilon=\beta_x=\beta_y=0.1$. The fractional Gaussian noises $\eta_x(t)$ and $\eta_y(t)$ were generated by following the Hosking method~\cite{Hosking}. Figure~\ref{fig1} shows examples of simulated trajectories for three different values of $h_y$ and keeping $h_x=0.5$. By visual inspection, we note that depending on whether the motion in $y$ is anti-persistent [Fig~\ref{fig1}(a)] or persistent [Fig~\ref{fig1}(c)], the diffusive behavior in the $x-$direction drastically changes. The anti-persistence in $y(t)$ causes the particle to return more often to the strip where the diffusive motion in the $x-$direction is allowed; consequently, $x(t)$ covers a larger interval along the $x-$axis when compared with the usual comb [$h_x=h_y=0.5$, Fig~\ref{fig1}(b)].
On the other hand, persistence in $y(t)$ causes the particle to return less often to the strip and, consequently, $x(t)$ covers a small interval along the $x-$axis.

\begin{figure}[!ht]
\centering
\includegraphics[scale=0.5]{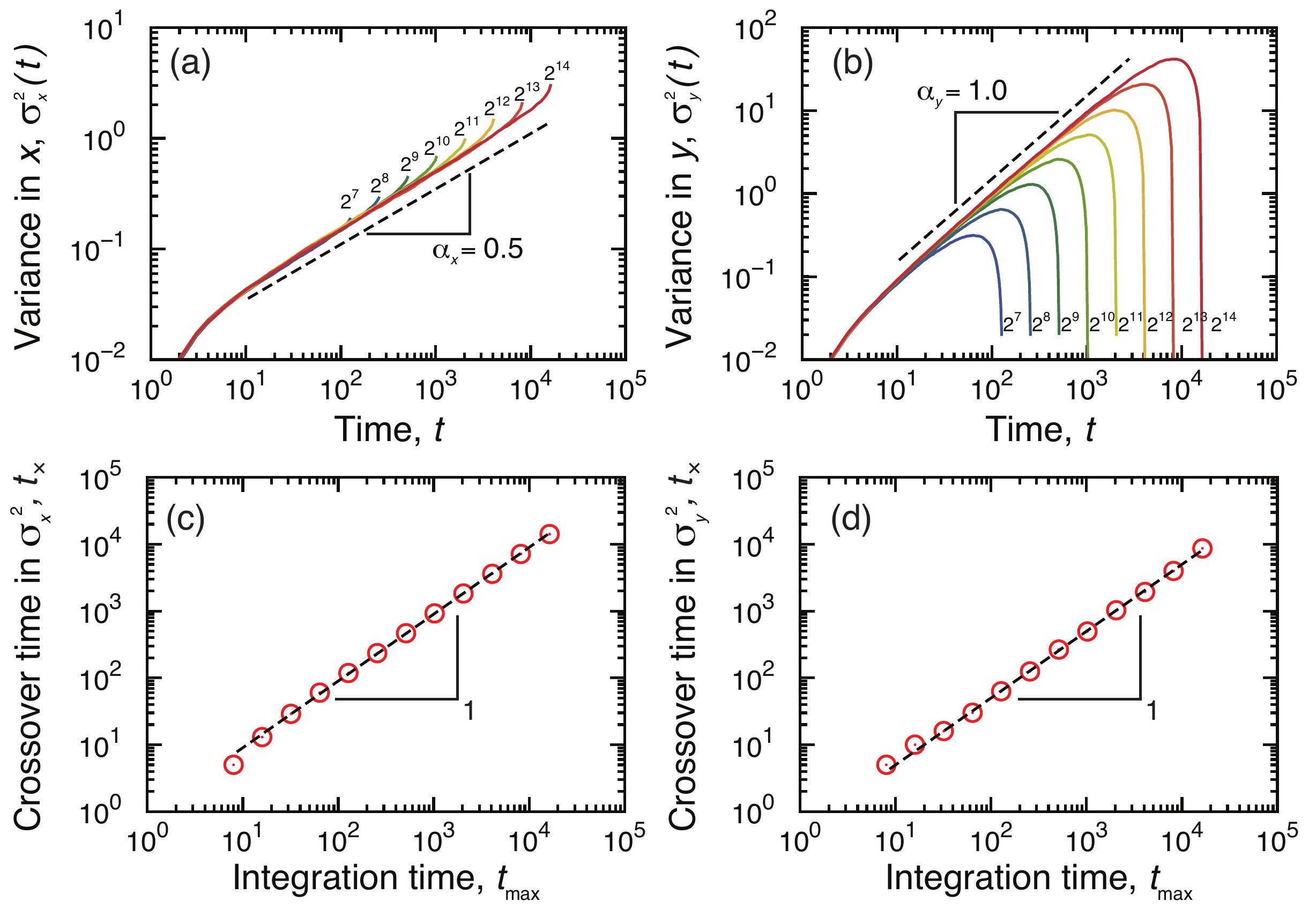}
\caption{(color online) Equivalence between the variances simulated via the Langevin equations and the analytical results obtained from the generalized diffusion equation in the usual case. Temporal dependence of the variances (a) $\sigma_x^2(t)$ in the $x-$direction and (b) $\sigma_y^2(t)$ in the $y-$direction for several values of the maximum integration time $t_{\text{max}}$ (as indicated in the plots). The dashed lines represent the predictions of the generalized diffusion equation, that is, $\sigma_x^2(t)\sim t^{\alpha_x}$ with $\alpha_x=0.5$ and $\sigma_y^2(t)\sim t^{\alpha_y}$ with $\alpha_y=1$~\cite{Arkhincheev2,Arkhincheev3,daSilva}. We note a finite-size-like effect in both plots. In $(c)$ and $(d)$, we show the crossover time $t_\times$, where the variances $\sigma_x^2(t)$ and $\sigma_y^2(t)$ start to change their behaviors, as a function of the maximum integration time $t_{\text{max}}$. We observe that $t_\times$ grows linearly with $t_{\text{max}}$ in both cases; therefore, in the limit as $t_{\text{max}}$ approaches infinity, the power-law regimes must prevail. 
}\label{fig2}
\end{figure}

Before quantifing the effects of the long-range correlations on the variances of $x(t)$ and $y(t)$, we have first verified whether the Langevin equations [Eq.~(\ref{leqcomb})] are actually equivalent to the generalized diffusion equation [Eq.~(\ref{eqcombusual})] when the noises $\eta_x(t)$ and $\eta_y(t)$ are uncorrelated ($h_x=h_y=0.5$). We thus simulate an ensemble of $10^5$ particle positions considering $dt=1$ and $h_x=h_y=0.5$. By using these data, we evaluate the temporal evolution of the variances 
\begin{eqnarray}
\sigma^2_x(t)&=&\langle( x(t) - \langle x(t) \rangle)^2\rangle~~\text{and}~~\nonumber \\
\sigma^2_y(t)&=&\langle( y(t) - \langle y(t) \rangle)^2\rangle\,,
\end{eqnarray}
where $\langle\dots\rangle$ denotes ensemble average. Figures~\ref{fig2}(a) and (b) show the behavior of $\sigma^2_x(t)$ and $\sigma^2_y(t)$ versus time for several values of the maximum integration time $t_{\text{max}}$ in the Langevin equations, ranging from $2^{7}$ to $2^{14}$ as indicated in these plots. In both cases, we  note a finite-size-like effect characterized by a crossover time $t_\times$ where $\sigma^2_x(t)$ and $\sigma^2_y(t)$ change their behaviors. Despite that and also discarding an initial transient regime ($t \lesssim10$), we observe that the variance profiles of $\sigma^2_x(t)$ and $\sigma^2_y(t)$ are in well agreement with the power laws predicted by the generalized diffusion equation of the usual comb [Eq.~(\ref{eqcombusual})], that is, $\sigma^2_x(t)\sim t^{\alpha_x}$ and $\sigma^2_y(t)\sim t^{\alpha_y}$ with $\alpha_x=0.5$ and $\alpha_y=1$~\cite{Arkhincheev2,Arkhincheev3,daSilva}. We have also analyzed the behavior of the crossover time $t_\times$ in function of the maximum integration time $t_{\text{max}}$. Figures~\ref{fig2}(c) and (d) show that $t_\times$ grows linearly with $t_{\text{max}}$ for $\sigma^2_x(t)$ and $\sigma^2_y(t)$, confirming that the simulated power-law regimes prevail when $t_{\text{max}}$ tends to infinity. 

\begin{figure}[!ht]
\centering
\includegraphics[scale=0.5]{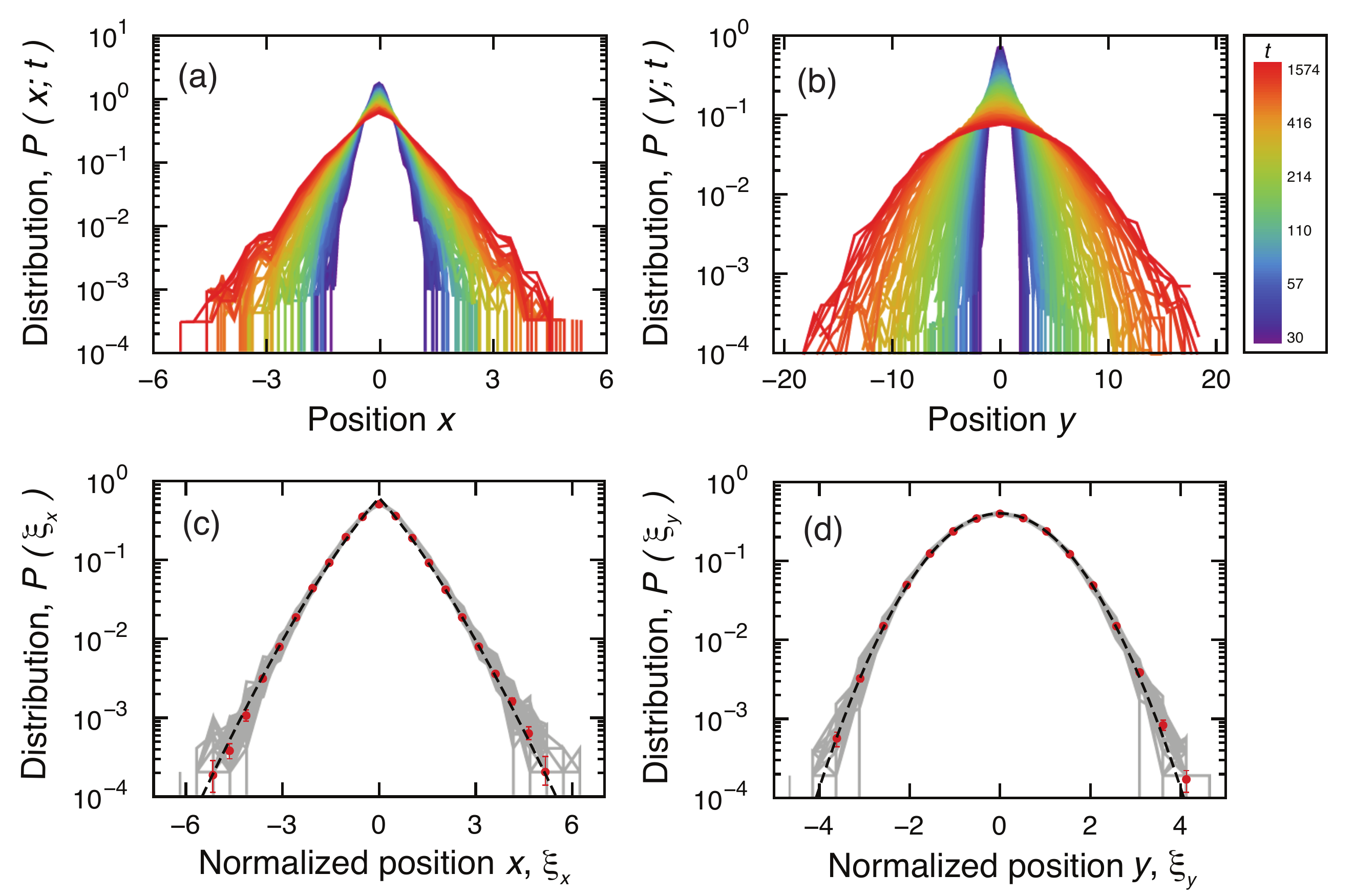}
\caption{(color online) Equivalence between the marginal distributions $P(x;t)$ and $P(y;t)$ simulated via the Langevin equations and the analytical results obtained from the generalized diffusion equation in the usual case. Marginal distributions (a) $P(x;t)$ and (b) $P(y;t)$ for several values of $t$, as indicated by color code. We note the spreading of these distributions and the normality of $P(y;t)$. Marginal distributions of normalized positions (c) $P(\xi_x)$ and (d) $P(\xi_y)$ for the same values of $t$ of the previous figures. The gray lines are the marginal distributions for each value of $t$ and the red circles are window average values over all distributions (the error bars are 95\% confidence intervals obtained via bootstrapping). The dashed lines are the analytical predictions of Eq.~(\ref{eqcombusual}); specifically, in (c) the dashed line is the distribution of Eq~(\ref{pxixcollapsed}) and in (d) it is the standard Gaussian $P(\xi_y)=(1/\sqrt{2 \pi}) \exp(-\xi_y^2/2)$. In addition to the good (visual) agreement between simulations and analytical expressions, we have found that the $p$-values of the Kolmogorov-Smirnov test (testing the equality of the empirical and analytical distributions) are all larger than $0.05$, indicating that the hypothesis of the analytical distributions describing our data cannot be rejected at a confidence level of 95\%.
}\label{fig3} 
\end{figure}

In order to further strengthen the equivalence between our Langevin equations~(\ref{leqcomb}) and the diffusion equation~(\ref{eqcombusual}), we have also evaluated the marginal probability functions \mbox{$P(x;t)= \int \rho(x,y;t) dy$} and \mbox{$P(y;t)= \int \rho(x,y;t) dx$} from the simulated data. Figures~\ref{fig3}(a) and~\ref{fig3}(b) show these distributions for several values of $t$ where we note the spreading of these distributions. For a better comparison with the analytical predictions of Eq.~(\ref{eqcombusual}), we have also calculated the marginal distributions considering the normalized positions $\xi_x=[x(t)-\langle x(t) \rangle]/\sigma_x(t)$ and $\xi_y=[y(t)-\langle y(t) \rangle]/\sigma_y(t)$. Figures~\ref{fig3}(c) and~\ref{fig3}(d) reveal a good collapse of the empirical distributions and also a good agreement with the analytical distributions (dashed lines) predicted by Eq.~(\ref{eqcombusual}), which is a standard Gaussian for $P(\xi_y)$ and can be expressed in terms of the Fox H function for $P(x;t)$~\cite{Arkhincheev2,Arkhincheev3,daSilva} or by the following summation~\cite{Metzler,Girst}
\begin{equation}\label{pxixcollapsed}
P(\xi_x) = \frac{1}{\pi^{1/4}} \sum_{n=0}^{\infty}\frac{(-1)^n}{n!\,\Gamma[1-(n+1)/4]}\left(\frac{\xi_x^2}{\sqrt{\pi}/4}\right)^{n/2}\,.
\end{equation}
Thus, for the case where no correlations are present in the noises $\eta_{x}(t)$ and $\eta_{y}(t)$, the results shown in Fig.~\ref{fig3} corroborate with the equivalence between our simulations of the Langevin equations and the diffusion equation for the comb model.

\begin{figure*}[!ht]
\centering
\includegraphics[scale=0.5]{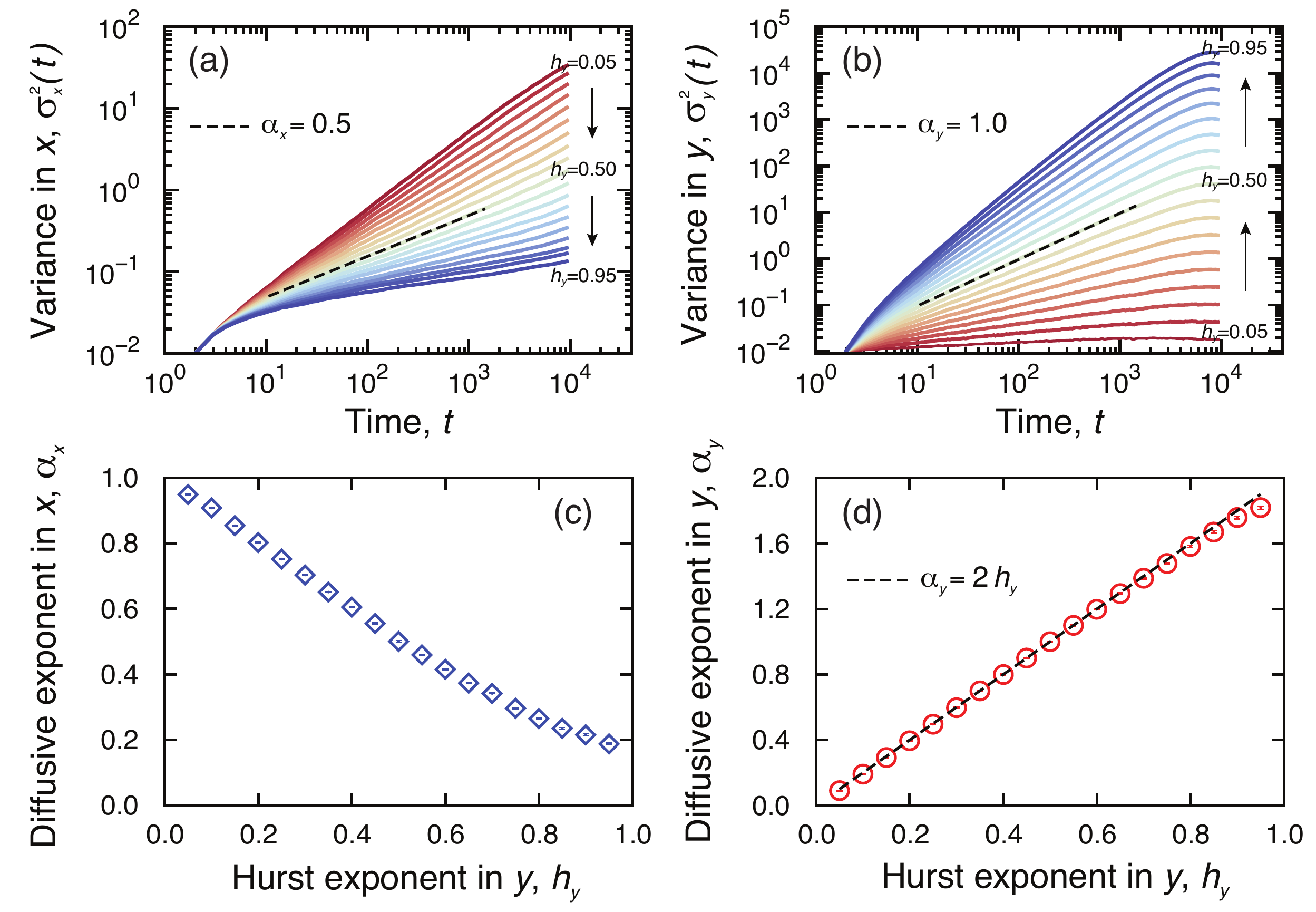}
\caption{(color online) The role of the long-range correlated noise in the $y-$direction on the variances $\sigma^2_x(t)$ and $\sigma^2_y(t)$. Profile of (a) $\sigma^2_x(t)$ and (b) $\sigma^2_y(t)$ for nineteen values of $h_y$ homogeneously distributed between $0.05$ and $0.95$. In both cases, the dashes lines represent the behavior of the usual comb:  $\sigma^2_x(t)\sim t^{\alpha_x}$ with $\alpha_x=0.5$ and $\sigma^2_y(t)\sim t^{\alpha_y}$ with $\alpha_y=1$. Here, we have omitted the finite-size-like effect on the variances for better visualization of the power-law behavior. For $\sigma^2_x(t)$, we observe that the diffusion is enhanced (when compared with the usual comb) for an anti-persistent noise in $y$ ($h_y<0.5$) and that is slows down when the noise in $y$ is persistent. In the case of $\sigma^2_y(t)$,  anti-persistent noise slows down the diffusion while persistent noise enhances it. In $(c)$ and $(d)$, we show the dependence of $\alpha_x$ and $\alpha_y$ on $h_y$. The values of $\alpha_x$ and $\alpha_y$ were obtained by least squares fitting a linear model to the log-log relationships between the variances and the time $t$. The error bars (which are very tiny) are the standard errors of $\alpha_x$ and $\alpha_y$. When adjusting the data, we have selected only the values $\sigma^2_x(t)$ and $\sigma^2_y(t)$ within the interval $30\leq t\leq1600$ in order to avoid the initial transient regime and the finite-size-like effect. We note the monotonic but nonlinear decay of $\alpha_x$ in function of $h_y$ and the straightforward behavior of $\alpha_y$ versus $h_y$ described by $\alpha_y=2 h_y$ (dashed line).}\label{fig4}
\end{figure*}

Convinced of the equivalence in the usual case, we now address the role of the long-range correlations on the diffusive properties of a random walker described by our Langevin equations~(\ref{leqcomb}). We first investigate how the memory effects in the $y-$direction modify the diffusion in the $x$ and $y$ directions. In order to do so, we have built an ensemble of $10^5$ particle positions considering $dt=1$ and $h_x=0.5$ for each one of nineteen values of $h_y$ homogeneously distributed between $0.05$ and $0.95$. For each value of $h_y$, we evaluate the temporal evolution of the variances $\sigma^2_x(t)$ and $\sigma^2_y(t)$. Figures~\ref{fig4}(a) and~\ref{fig4}(b) show the variance profiles in the $x$ and $y$ directions, respectively. In these plots, the dashed lines represent the behavior for the usual comb, that is, when no correlations are present. In the case of $\sigma^2_x(t)$, we note that the more the noise in $y$ is anti-persistent ($h_y<0.5$), the larger is the power-law exponent $\alpha_x$ describing the behavior of $\sigma^2_x(t)$; in contrast, persistent noise in $y$ slows down the diffusion and decreases the power-law exponent $\alpha_x$. The extreme cases where $h_y=0.05$ and $h_y=0.95$ are characterized by $\alpha_x\approx0.95$ and  $\alpha_x\approx0.19$, respectively. Thus, as we have first noticed in Fig.~\ref{fig1}, anti-persistent noise in $y$ forces the walker to return more frequently to the strip where it can move in the $x-$direction and, consequently, the diffusion is enhanced when compared with the case usual comb ($h_y=0.5$). 
On the other hand, persistent noise in $y$ drives the walker away from strip and, consequently, the diffusion slows down. In the case of $\sigma^2_y(t)$, the behavior is straightforward: anti-persistent noise in $y$ slows down the diffusion while persistent noise enhances it [see Fig.~\ref{fig4}(b)]. In order to quantify the role of long-range correlations in $y$ on $\sigma^2_x(t)$ and $\sigma^2_y(t)$, we have evaluated the dependence of the power-law exponents $\alpha_x$ and $\alpha_y$ on $h_y$. In Fig.~\ref{fig4}(c), we observe a monotonic but nonlinear decay of $\alpha_x$ as $h_y$ increases, while in Fig.~\ref{fig4}(d), we confirm the straightforward behavior of $\alpha_y$ as function of $h_y$ ($\alpha_y=2 h_y$).

\begin{figure}[!ht]
\centering
\includegraphics[scale=0.5]{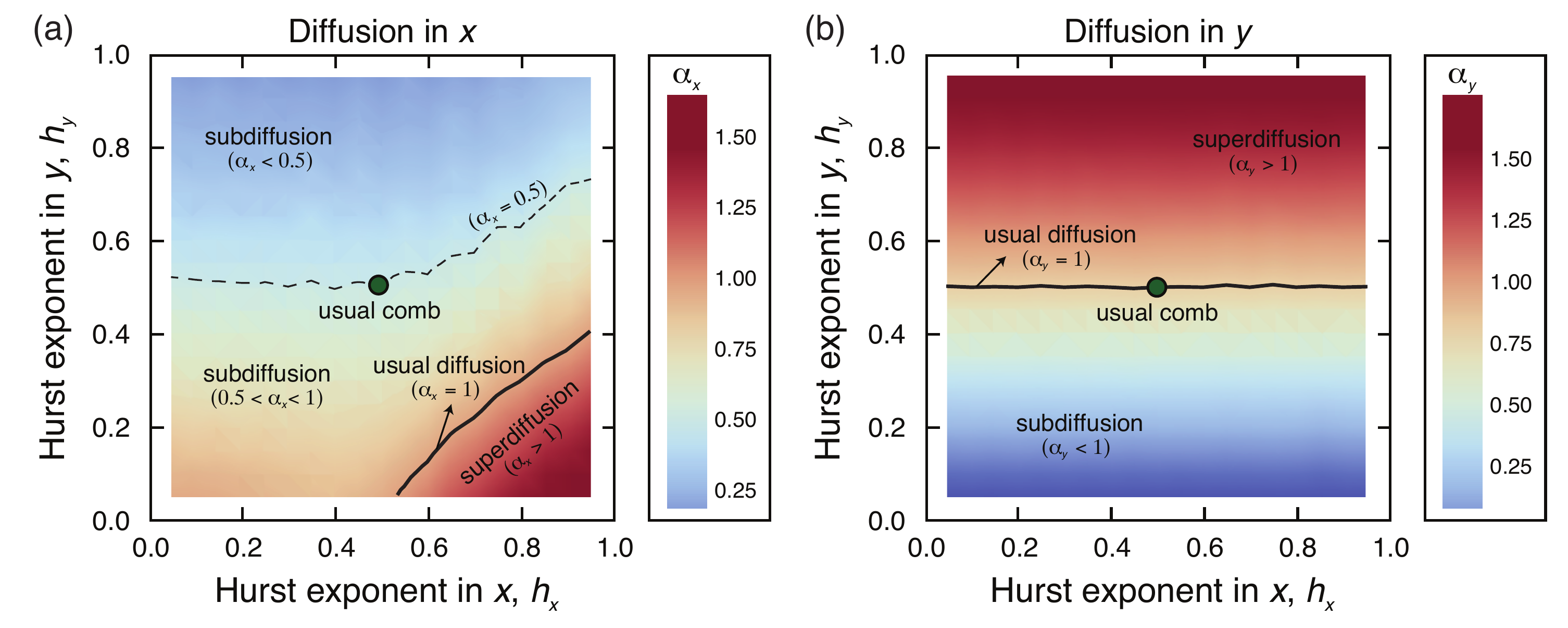}
\caption{(color online) The interplay between the long-range correlated noises in the $x$ and $y$ directions on the variances $\sigma^2_x(t)$ and $\sigma^2_y(t)$. Contour plots of (a) $\alpha_x$ and (b) $\alpha_y$ as a function of $h_x$ and $h_y$. We note the rich diffusive scenery in the $x-$direction where we may have subdiffusion slower ($\alpha_x<0.5$) or faster ($0.5<\alpha_x<1$) than the usual comb, usual diffusion ($\alpha_x=1$) and even superdiffusion ($\alpha_x>1$). We further observe that the power-law exponent $\alpha_x$ cannot be expressed as a linear combination of $h_x$ and $h_y$, suggesting a more complex coupling between the Hurst exponents and diffusion in the $x-$direction.}\label{fig5}
\end{figure}

Another interesting question is whether the combinations of long-range correlated noises in the $x$ and $y$ directions affect the diffusive behaviors in a nontrivial way. To answer this question, we have built an ensemble of $10^5$ particle positions for each possible pair of $h_x$ and $h_y$ homogeneously distributed between $0.05$ and $0.95$ with step size of $0.05$. For each combination of $h_x$ and $h_y$, we calculate the profile of the variances $\sigma^2_x(t)$ and $\sigma^2_y(t)$ and also the power-law exponents $\alpha_x$ and $\alpha_y$ describing the main tendency of these curves (by using the same procedure of Fig.~\ref{fig4}). Figure~\ref{fig5}(a) shows a contour plot of $\alpha_x$ as a function of $h_x$ and $h_y$ and Fig.~\ref{fig5}(b) represents the same for $\alpha_y$. For the Fig.~\ref{fig5}(a), we note a quite rich diffusive scenery where, depending on the values of $h_x$ and $h_y$, we may have subdiffusion slower ($\alpha_x<0.5$) or faster ($0.5<\alpha_x<1$) than the usual comb, usual diffusion ($\alpha_x=1$) and even superdiffusion ($\alpha_x>1$). We further observe from this figure that $\alpha_x$ cannot be written as a linear combination of $h_x$ and $h_y$ (which would be represented by straight lines in this plot) and thus, the correlations in $x$ and $y$ present a nontrivial interplay in the behavior of $\alpha_x$. On the other hand, the case of Fig~\ref{fig5}(b) is rather simple because the correlations in $x$ do not affect the diffusive behavior in the $y-$direction, this becomes evident by noting that the level curves of $\alpha_y$ are horizontal straight lines.

\begin{figure}[!ht]
\centering
\includegraphics[scale=0.5]{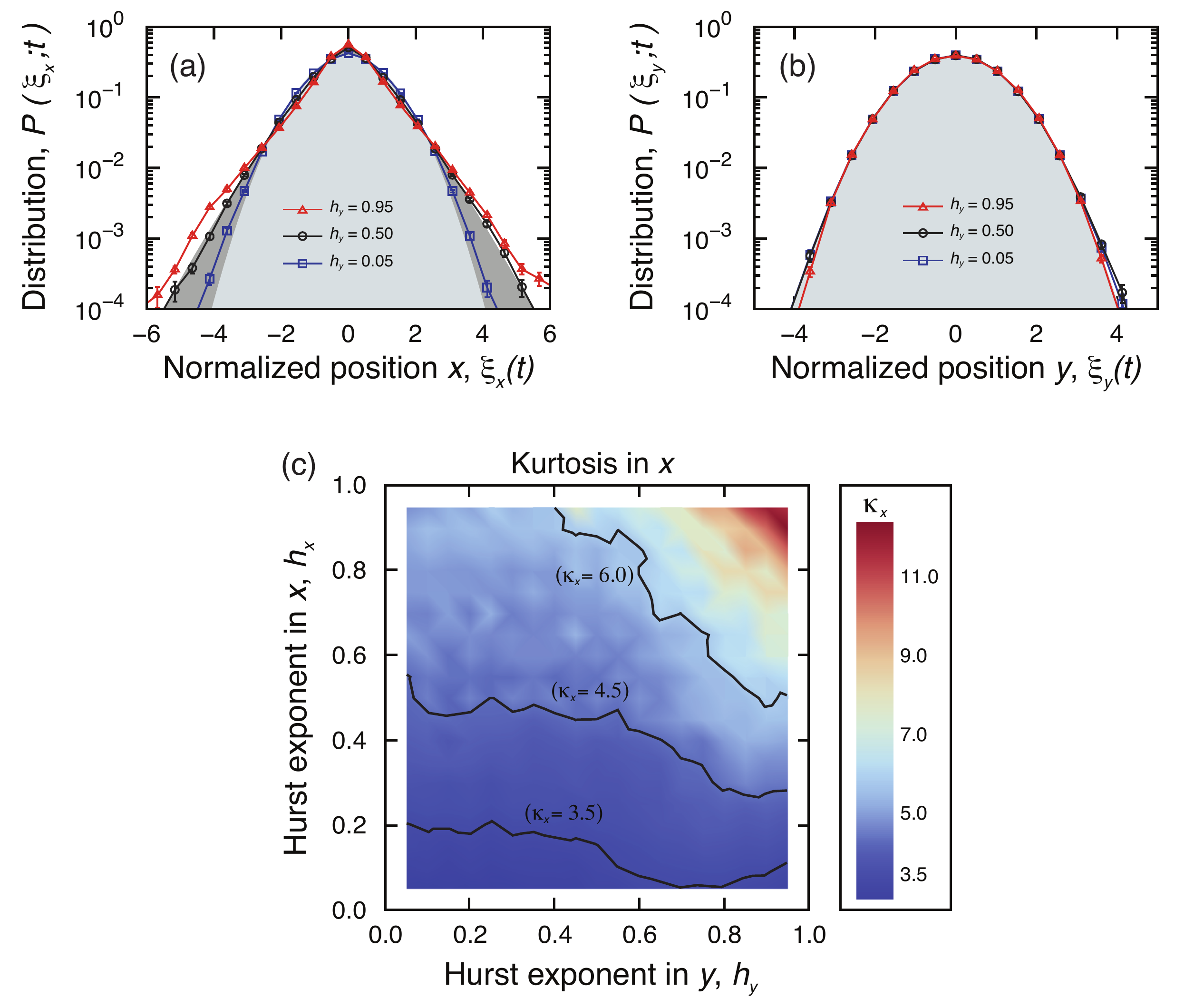}
\caption{(color online) The effect of the long-range correlated noises on the marginal distributions. Marginal distributions of normalized positions (a) $P(\xi_x)$ and (b) $P(\xi_y)$ for there values of $h_y$ (indicated in the plots) and $h_x=0.5$. In both plots, the light shaded regions represent a standard Gaussian, while the dark shaded region in (a) is the profile of the marginal distribution for $\xi_x$ of the usual comb [Eq.~(\ref{pxixcollapsed})]. For $P(\xi_x)$, we note that the distribution tails decay slower than those of usual comb when the noise in the $y-$direction is persistent ($h_y=0.95$); whereas for anti-persistent noise ($h_y=0.05$), the tails decay faster and approach the Gaussian distribution. In the case $P(\xi_y)$, the long-range correlations in $y$ have practically no effect on the profile of these distributions. (c) Contour plot of the coefficient of kurtosis $\kappa_x=\langle \xi_x^4\rangle/\langle \xi_x^2\rangle^2$ for the normalized position $\xi_x$ as a function of $h_x$ and $h_y$. The value of $\kappa_x$ is an average over $t$ within the interval $30\leq t\leq1600$. We note that, depending on the values of $h_x$ and $h_y$, the distribution $P(\xi_x)$ exhibits different profiles, which can be mesokurtic ($\kappa_x=3$, as in the Gaussian case), leptokurtic ($\kappa_x>3$) or platykurtic ($\kappa_x<3$). We also observe that, similarly to variance case, $\kappa_x$ cannot be expressed in terms of linear combination of $h_x$ and $h_y$.}\label{fig6}
\end{figure}

It is also important to investigate whether the long-range correlations affect the distribution profiles of the particle positions. In order to do so, we first calculate the marginal distributions $P(\xi_x)$ and $P(\xi_y)$ for there different values of $h_y$ ($0.05$, $0.50$ and $0.95$) and keeping $h_x=0.50$. Figures~\ref{fig6}(a) and~\ref{fig6}(b) show these distributions. We note that the distribution $P(\xi_x)$ displays tails longer than those of usual comb (dark shaded region) when the noise in the $y-$direction is persistent ($h_y=0.95$) whereas, for anti-persistent noise ($h_y=0.05$), the tails are shorter than those of usual comb and close to the Gaussian distribution (light shaded region). The distributions $P(\xi_y)$ of the normalized positions $\xi_y$ are not affected by long-range correlated noises in the $y-$direction. We thus confirm that the long-range correlations in $y$ have influence on the profiles of $P(\xi_x)$. For a better and quantitative characterization of the role of the correlated noise $\eta_y(t)$ on the shape of $P(\xi_x)$, we evaluate the coefficient of kurtosis $\kappa_x=\langle \xi_x^4\rangle/\langle \xi_x^2\rangle^2$ in function of $h_x$ and $h_y$. For $\kappa_x>3$ the distribution is leptokurtic (peaked) while for $\kappa_x<3$ the distribution is platykurtic (flat) also, $\kappa_x=3$ for a Gaussian distribution and $\kappa_x\approx4.71$ for the usual comb (distribution of the Eq.~\ref{pxixcollapsed}). Figure~\ref{fig6}(c) shows a contour plot of $\kappa_x$ as a function of $h_x$ and $h_y$, where we observe that the kurtosis ranges from very close to $3$ ($\kappa_x=3.04\pm0.04$ for $h_x=h_y=0.05$) up to almost $13$ ($\kappa_x=12.55\pm0.87$ for $h_x=h_y=0.95$). It is worth noting that, similar to the variance case [Fig.~\ref{fig5}(a)], the kurtosis $\kappa_x$ is not a trivial combination of $h_x$ and $h_y$.

\section{Summary and Conclusions}
We have proposed an extension for the comb model via Langevin equations driven by long-range correlated noises (fractional Gaussian noises). We initially showed that our Langevin equations are equivalent to the generalized diffusion equation (describing the usual comb) through the comparison of the numerical-obtained variances and marginal distributions with the analytical results predicted by the diffusion equation. Next, we have presented an extensive characterization of the diffusive properties of the particle positions in function of the Hurst exponents $h_x$ and $h_y$ defining the correlated noises. Our results show that the noise in the $y-$direction affects the variances and marginal distributions along the $x-$direction. Specifically, long-range persistence in $y$ slows down the diffusion in $x$ by reducing the power-law exponent $\alpha_x$ and makes the tails of the marginal distribution $P(\xi_x)$ longer as $h_y>0.5$ increases; whereas long-range anti-persistence in $y$ enhances the diffusion in $x$ by increasing the value of $\alpha_x$ and makes the tails of $P(\xi_x)$ shorter as $h_y<0.5$ decreases. We have also investigated the interplay between the long-range correlations in the $x$ and $y$ directions, where we found that both $\alpha_x$ and the kurtosis coefficient $\kappa_x$ of $P(\xi_x)$ are not a trivial (such as a linear combination) function of $h_x$ and $h_y$, but instead our results suggest a more complex coupling between the Hurst exponents and the diffusion in the $x-$direction. In summary, a quite rich diffusive scenery emerges from our model which may be directly applied for describing (at least in a first approximation) some of the recent empirical findings related to diffusion in complex systems such as living cells. Furthermore, our model combines and allows the study/analysis of the interplay between different mechanisms of anomalous diffusion, geometric constraints and long-range correlations, an important fact that had not been explored within the comb model framework and that may provide a better understanding of these processes as a combination of different mechanisms.

\section*{Acknowledgments}
We are all grateful to Capes, CNPq and Funda\c{c}\~ao Arauc\'aria for financial support.

\end{document}